\documentclass[draft,jgrga]{agu2001}
\usepackage{graphicx}
\usepackage{lineno}
\linenumbers
\authorrunninghead{V\"{O}R\"{O}S ET AL.}
\titlerunninghead{TURBULENCE AND FLUCTUATIONS NEAR VENUS}
\authoraddr{Z. V\"{o}r\"{o}s,
Institute of Astro- and Particle Physics, University of Innsbruck, Technikerstr. 25, 6020 Innsbruck, Austria.
(zoltan.voeroes@uibk.ac.at)}

\begin{document}

%% ------------------------------------------------------------------------ %%
%
%  ENABLE IMAGE DISPLAY WHILE USING DRAFT MODE
%
%% ------------------------------------------------------------------------ %%
%
% Uncomment the following code (as well as \usepackage{graphicx} above)
% if you need to include images in draft mode
%\setkeys{Gin}{draft=false}
%
% PLEASE NOTE: WHEN YOU SUBMIT YOUR LATEX FILE TO GEMS, COMMENT OUT ANY COMMANDS
% THAT INCLUDE GRAPHICS.
% (See FIGURES section near the end of the file)
%

%% ------------------------------------------------------------------------ %%
%
%  TITLE
%
%% ------------------------------------------------------------------------ %%

\title{Intermittent turbulence, noisy fluctuations and wavy structures in the Venusian magnetosheath and wake}
%

%% ------------------------------------------------------------------------ %%
%
%  AUTHORS AND AFFILIATIONS - 3 methods
%
%% ------------------------------------------------------------------------ %%

% Method 1 (for all journals, except Reviews of Geophysics, which
% should use method 3):
% For three or fewer author/affiliation blocks, use \author{} and \affil{}

%\author{Z. V\"{o}r\"{o}s}
%\affil{Space Research Institute, Austrian Academy of Sciences, Graz, Austria}
% ---------------
% Method 2 (for all journals, except Reviews of Geophysics, which
% should use method 3):
% For more than three author/affiliation blocks,
% use \author{\altaffilmark{}} and \altaffiltext{}
% \altaffilmark will produce footnote;
% matching altaffiltext will appear at bottom of page.
% May use \\ to start a new line.
\smallskip
\smallskip
\smallskip
\authors{Z. V\"or\"os, \altaffilmark{1}
 T. L. Zhang, \altaffilmark{2}
 M. P. Leubner, \altaffilmark{1}
 M. Volwerk, \altaffilmark{2}
 M. Delva, \altaffilmark{2}
and W. Baumjohann, \altaffilmark{2}
 }

 \altaffiltext{1}
 {Institute of Astro- and Particle Physics, University of Innsbruck, Innsbruck, Austria.}
 \altaffiltext{2}{Space Research Institute, Austrian Academy of Sciences, Graz, Austria.}
%
 %\altaffiltext{3}{State Key Laboratory of Space Weather, Chinese Academy of Sciences, China.}
%
 %\altaffiltext{3}{Institute of Experimental Physics, Slovakia Academy of Sciences, Kosice, Slovak Republik.}
%  ABSTRACT
%
%% ------------------------------------------------------------------------ %%

% Do NOT include any \begin...\end commands within
% the body of the abstract.

\begin{abstract}
Recent research has shown that distinct physical regions in the Venusian induced magnetosphere are recognizable from the variations of
strength of the magnetic field and its wave/fluctuation activity. In this paper the statistical properties of magnetic fluctuations are
investigated in the Venusian magnetosheath and wake regions. The main goal is to identify the characteristic scaling features of
fluctuations along Venus Express (VEX) trajectory and to understand the specific circumstances of the occurrence of different types of scalings.
For the latter task we also use the results of measurements from the previous missions to Venus.
Our main result is that the changing character of physical interactions between the solar wind and the planetary obstacle is
leading to different types of spectral scaling in the near-Venusian space. Noisy fluctuations are observed in the magnetosheath, wavy structures
near the terminator and in the nightside near-planet wake. Multi-scale turbulence is observed at the magnetosheath boundary
layer and near the quasi-parallel bow shock. Magnetosheath boundary layer turbulence is associated with an average magnetic field
which is nearly aligned with the Sun-Venus line. Noisy magnetic fluctuations are well described with the Gaussian statistics.
Both magnetosheath boundary layer and near shock turbulence statistics exhibit non-Gaussian features and intermittency over
small spatio-temporal scales. The occurrence of turbulence near magnetosheath boundaries can be responsible for the local heating of plasma
observed by previous missions.

\end{abstract}

%% ------------------------------------------------------------------------ %%
%
%  BEGIN ARTICLE
%
%% ------------------------------------------------------------------------ %%

% The body of the article must start with a \begin{article} command,
% and an \end{article} command must be placed at the end of the file,
% before \end{document}.
%
% If using draft mode \end{article} must follow the references section.

\begin{article}

%% ------------------------------------------------------------------------ %%
%
%  TEXT
%
%% ------------------------------------------------------------------------ %%
 \begin{figure*}
\noindent\includegraphics[width=40pc]{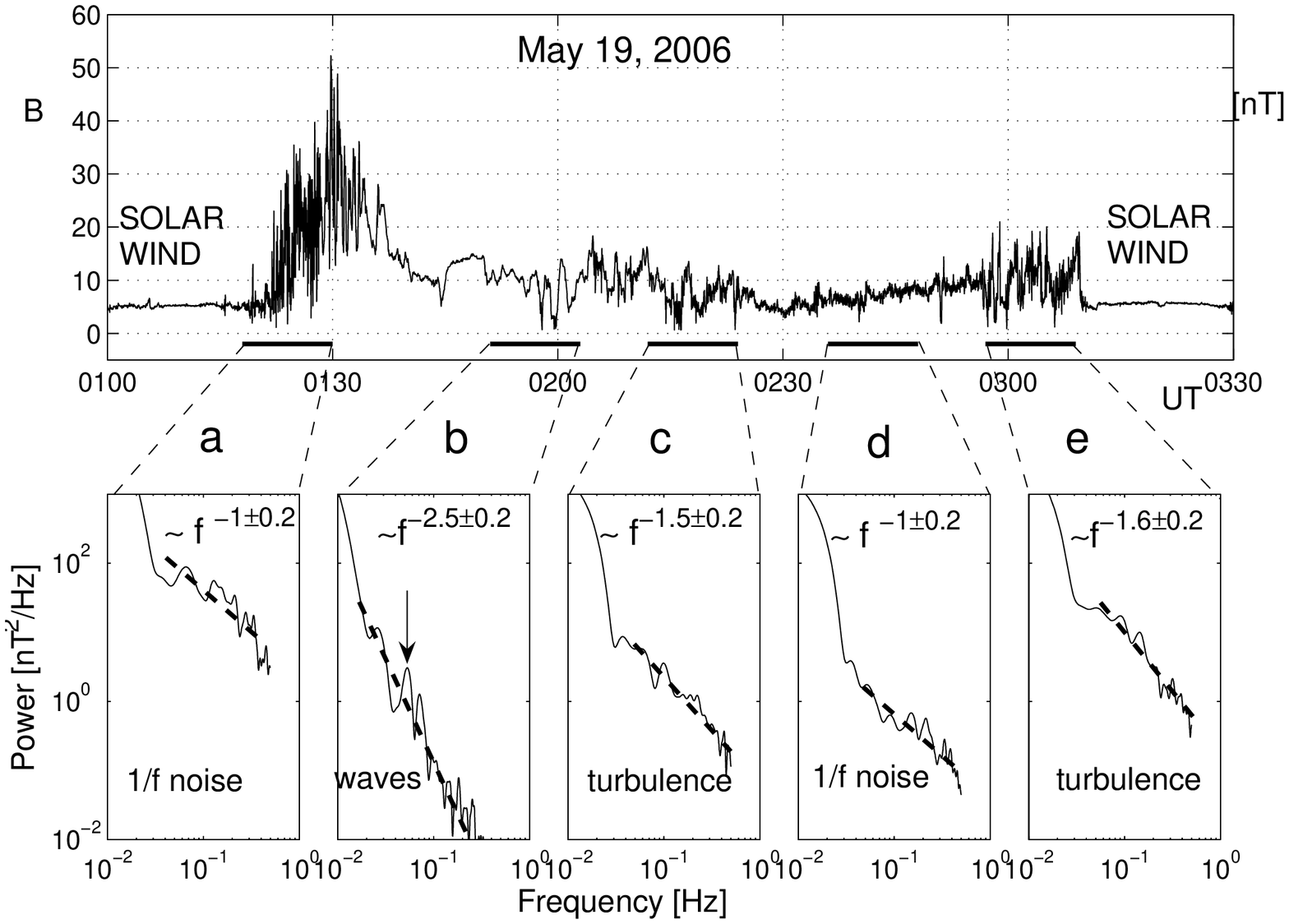}
 \caption{Top: Magnetic field strength (B) on May 19, 2006. The horizontal black lines correspond to the time intervals \textbf{a} - \textbf{e} of equal length in the dayside magnetosheath (\textbf{a}), night side near-planet wake (\textbf{b}), magnetosheath boundary layer (\textbf{c}), tailward magnetosheath (\textbf{d}), and in the vicinity of the bow shock; Bottom: Power spectra and spectral scalings estimated within the intervals \textbf{a} - \textbf{e}. Pronounced wavy structures are present mainly in the wake (interval \textbf{b}). The vertical arrow points to the spectral peak at $\sim$ 15 s.}
 \end{figure*}

\begin{figure*}
\noindent\includegraphics[width=45pc]{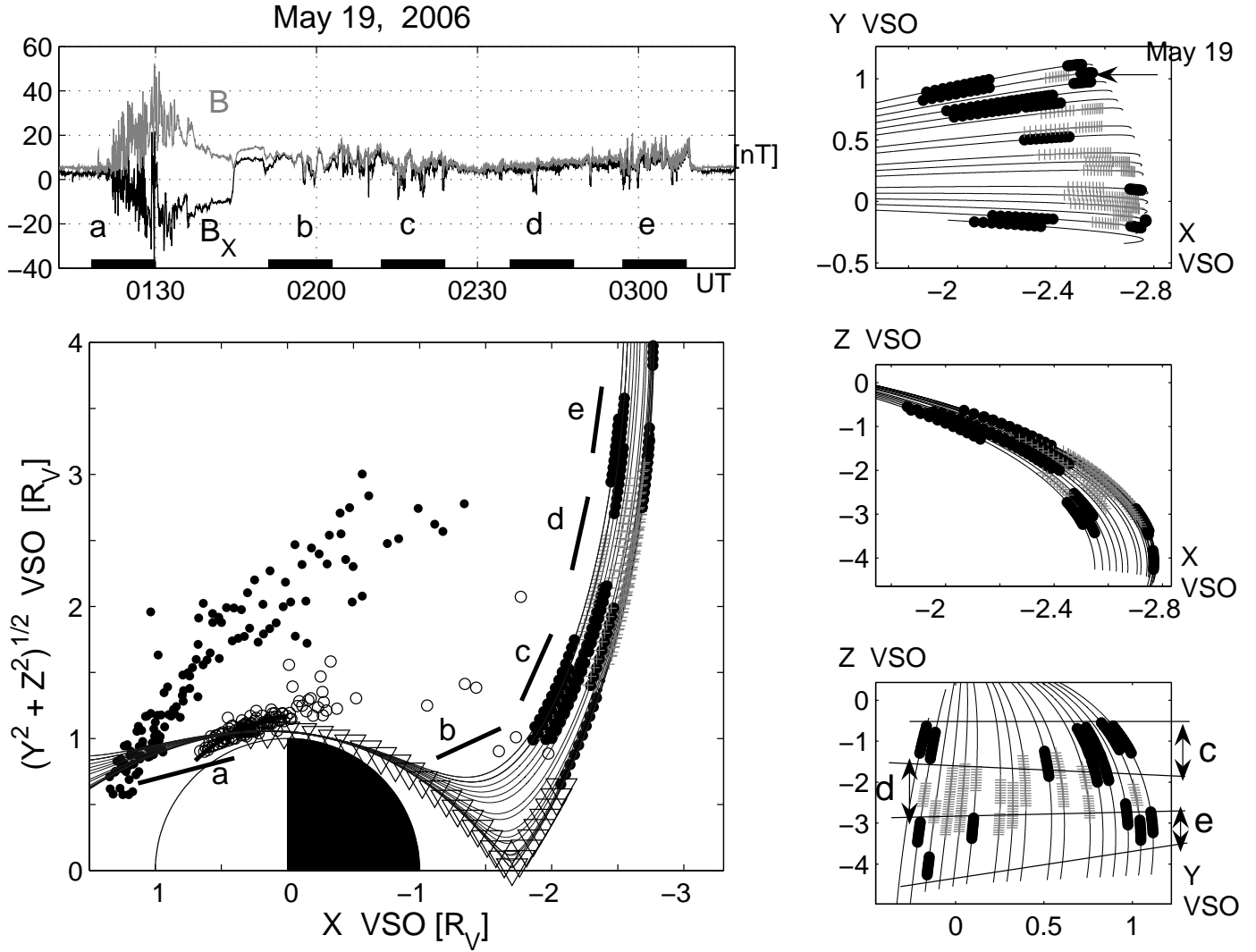}
\caption{Top left: Magnetic field strength (B, grey line) and $B_X$ magnetic component (black line) on May 19, 2006; the horizontal lines
show the intervals \textbf{a}-\textbf{e} depicted in Figure 1; Bottom left: VEX crossings (thin black lines) of the Venusian plasma environment in VSO coordinates. The intervals \textbf{a}-\textbf{e} (black lines) are shown alongside the VEX trajectories. The filled circles show the bow shock, the open circles show the magnetopause during multiple crossings (after Zhang et al., 2007). The large triangles along the lowest trajectory correspond to the region of wavy structures and spectral scaling index $\alpha \sim 2.5$. The family of thick black lines along the trajectories (intervals \textbf{c} and \textbf{e}) correspond to turbulent regions with $\alpha \sim 1.6$. The grey '+' signs are intervals of $1/f$ noise (interval \textbf{d});
Right: Enlarged magnetosheath crossings in VSO coordinates. The same notations is used as in the left. The bottom right subplot shows the approximate spatial size of the turbulent (\textbf{c, e}) and noisy (\textbf{d})regions in space.}
 \end{figure*}

 \begin{figure*}
\noindent\includegraphics[width=45pc]{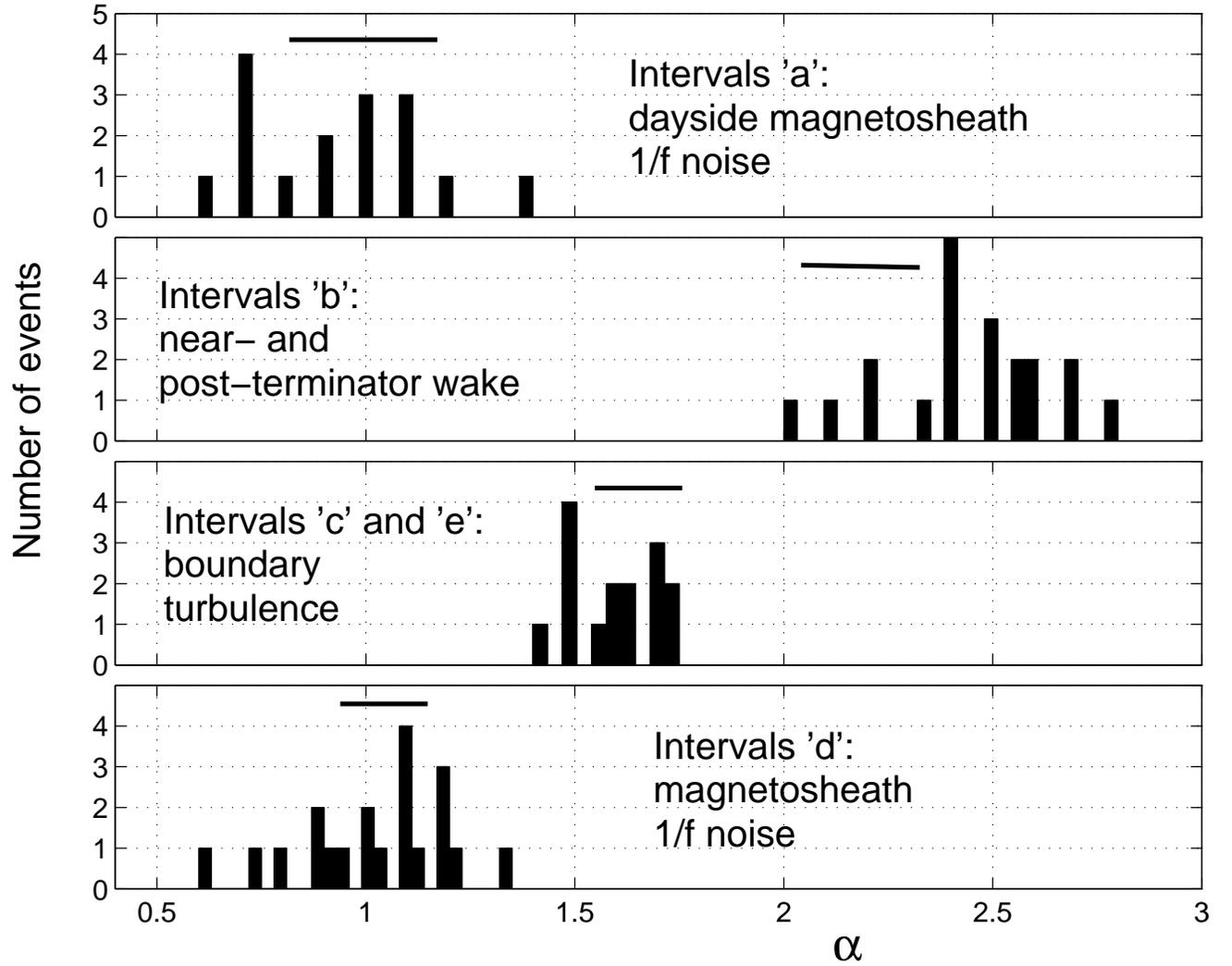}
 \caption{Histograms of the scaling indices in different regions of near-Venusian space (approximately obtained during the intervals \textbf{a-e} in Figure 2.). From top to bottom: dayside magnetosheath (\textbf{a}); near and post-terminator wake (\textbf{b}); magnetosheath boundary and near bow shock region (\textbf{c-e}); post-terminator tailward magnetosheath (\textbf{d}). These spatial regions are depicted more clearly in Figure 4. }
 \end{figure*}

 \begin{figure*}
\noindent\includegraphics[width=45pc]{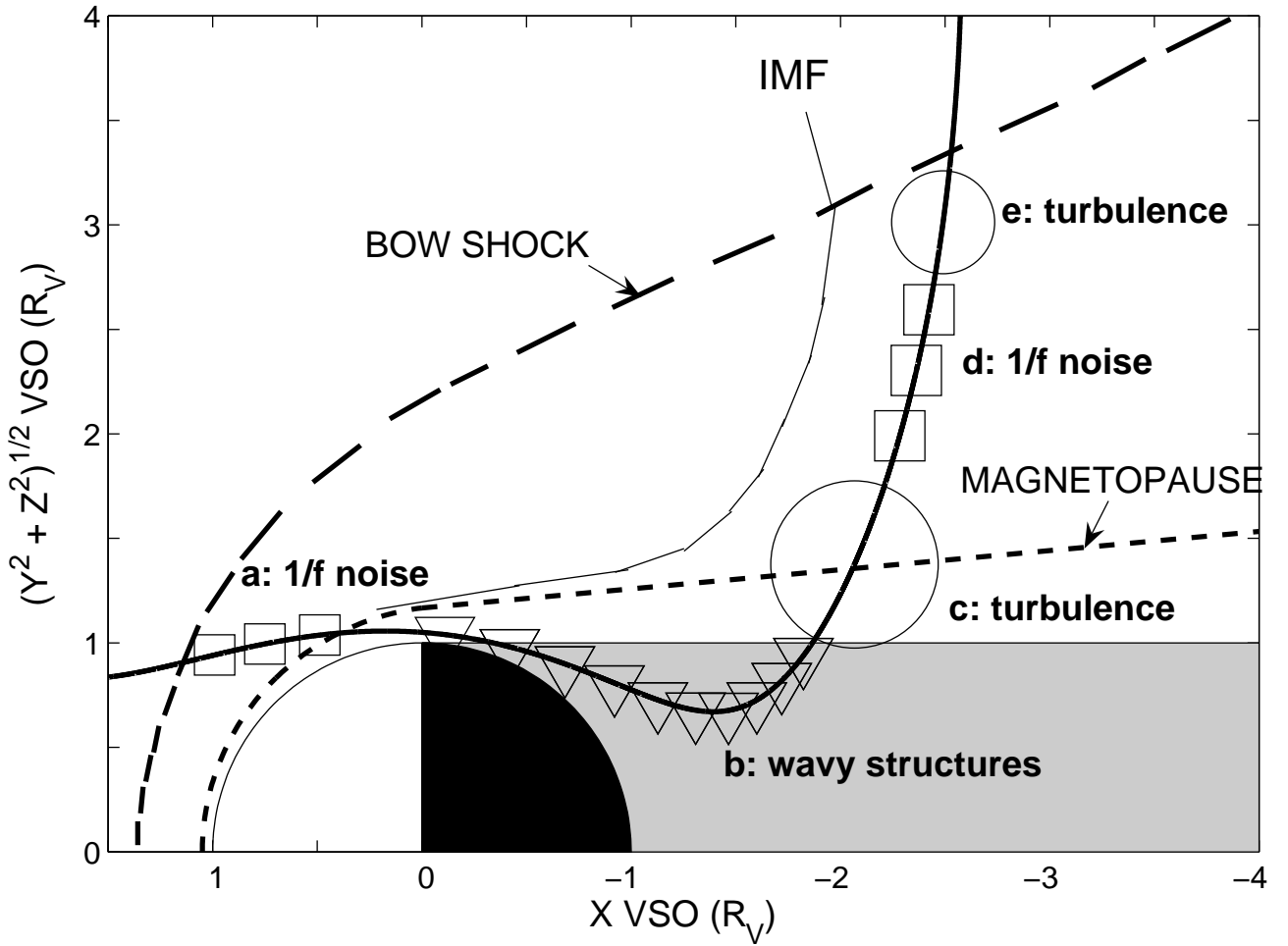}
 \caption{A cartoon showing the type of physical processes in the spatial regions of near-Venusian space; the magnetopause and the bow shock
 (dashed lines); a VEX crossing (solid black line); the optical shadow (shaded region); draped IMF (thin black line); the time intervals/spatial regions
 \textbf{a-e} are marked with squares, triangles, circle, squares, and circle, respectively.  }
 \end{figure*}

 \begin{figure}
\noindent\includegraphics[width=40pc]{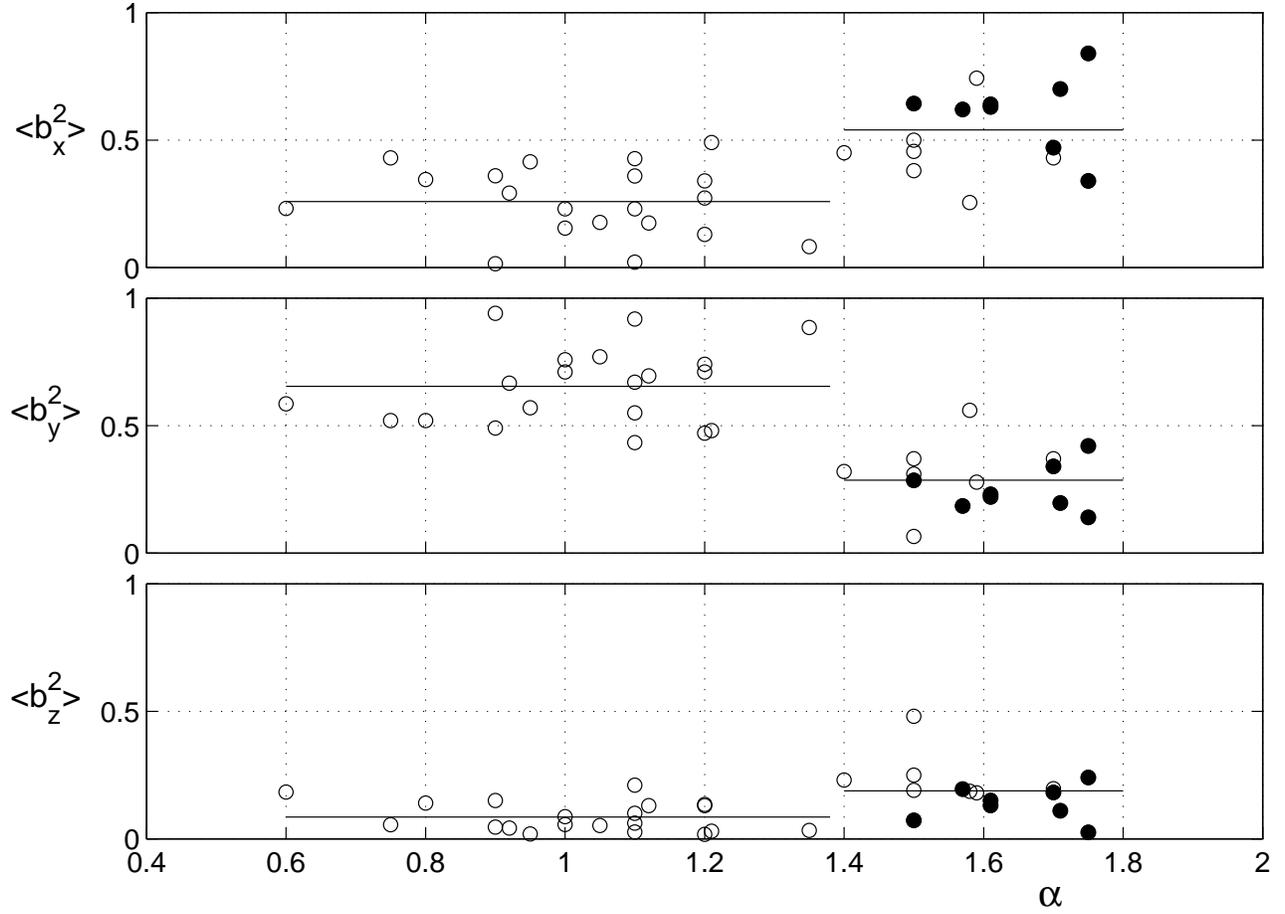}
 \caption{From top to bottom: comparison of 12 min averages of $<b_{X}^2>$,  $< b_{Y}^2>$ and
$<b_{Z}^2>$ with the observed spectral scaling indices $\alpha$;
 The subscripts indicate VSO magnetic field components;
 The horizontal lines correspond to averages of points (open and filled circles) over the  noisy $\alpha \in (0.6-1.4)$
 and turbulent $\alpha = 1.6 \pm 0.2$ scaling index ranges in each subplot; The turbulent intervals include boundary layer (filled circles)
 and near-shock (open circles) events;
 For each 12 min interval $b_{X}^2(t)+b_{Y}^2(t)+ b_{Z}^2(t) = 1$.
 }
 \end{figure}

 \begin{figure}
\noindent\includegraphics[width=45pc]{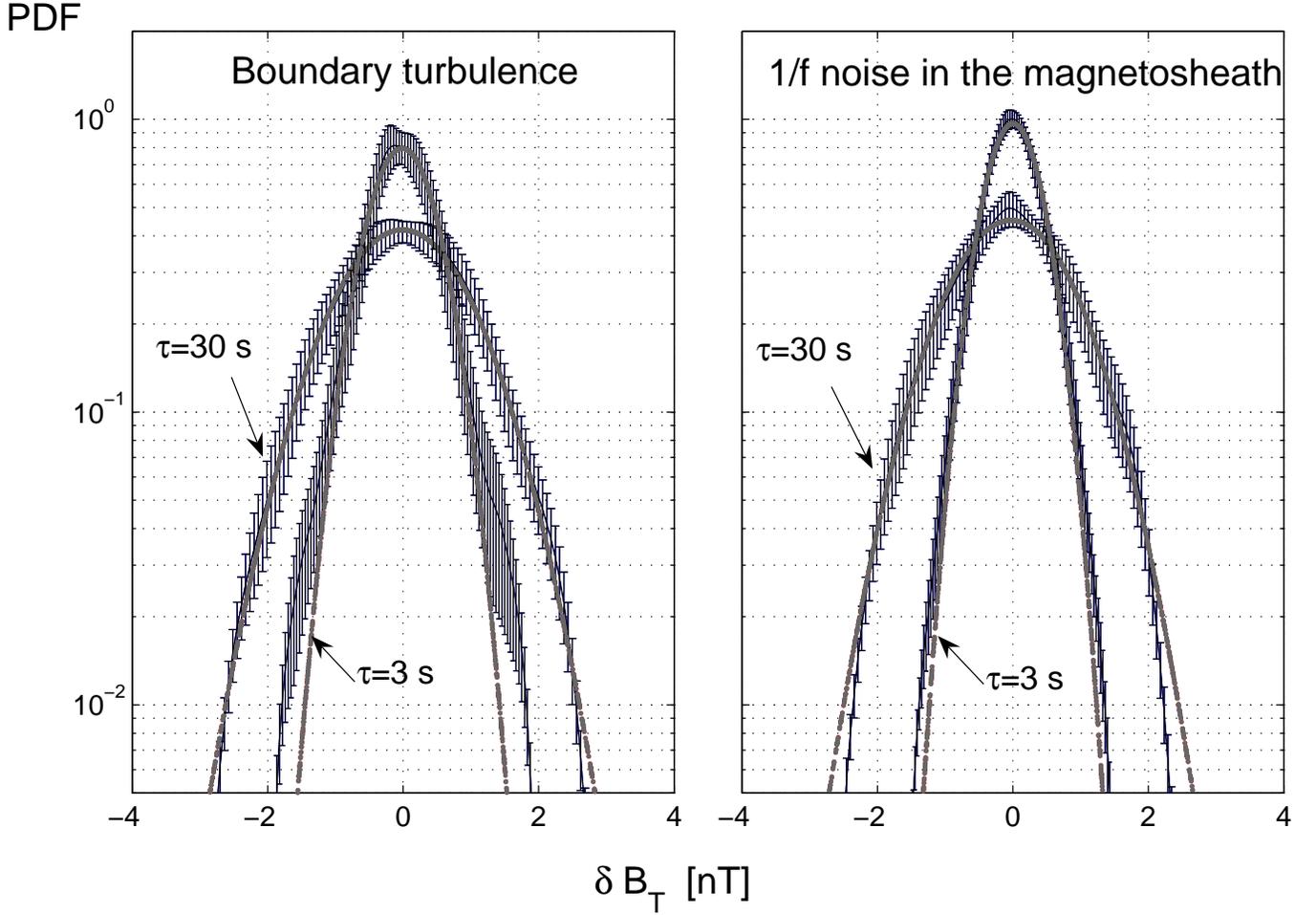}
 \caption{
Probability density functions (PDFs) constructed from magnetic time series (two-point differences defined through $\delta B = B(t+\tau) - B(t)$ )
of turbulent (left) and noisy (right) intervals. Gaussian fits are shown as dashed grey lines. At small scales (e.g. $\tau = 3$ s) the PDF is a non-Gaussian for
turbulent time series and Gaussian for noisy time series. At large scales (e.g. $\tau = 30$ s) the PDF is a Gaussian in both cases.
The error bars represent 95\% confidence limits for the mean in each point of $\delta B_T$.}
 \end{figure}

 \begin{figure}
\noindent\includegraphics[width=45pc]{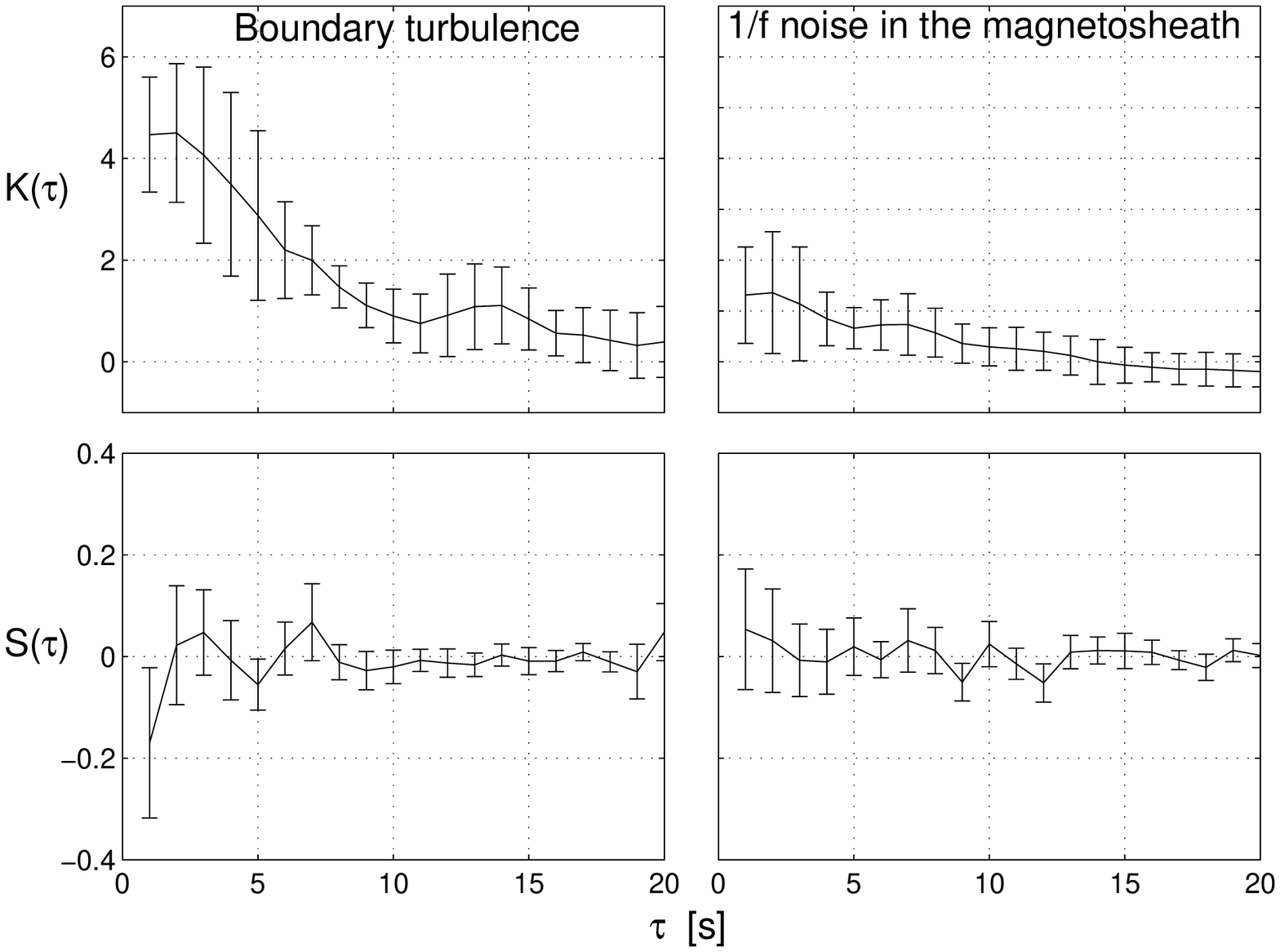}
 \caption{ The evolution of kurtosis ($K$, top) and skewness ($S$, bottom) with the time scale $\tau$, computed
 from two-point differences defined through $\delta B = B(t+\tau) - B(t))$ of turbulent (left) and noisy (right) time intervals.
 $K$ - peakedness of the PDF; $S$ - asymmetry around the mean, both relative to a Gaussian distribution,
for which $S=K=0$; The error bars represent 95\% confidence limits for the mean in each value of $\tau$.}
 \end{figure}

\section{Introduction}
The structured plasma environment of Venus is a natural plasma laboratory  where
the planetary ionosphere acts as an  obstacle to the supersonic solar wind  flow
carrying a magnetic  field. The absence  of an intrinsic  magnetic field ensures
that this interaction is more comet-like  than Earth-like. It is known from  the
previous missions  to Venus  (e.g., Russell  and Vaisberg,  1983) that  the most
prominent features include the direct interaction of ionized magnetosheath  flow
with the ionosphere/quasi-neutral atmosphere, mass loading of the  magnetosheath
flux-tubes, and the  transport/convection of magnetic  flux to the  wake region,
representing also the dominant source  for the magnetotail fluxes. However,  for
most of the time, mainly during time intervals of low solar wind dynamic pressure,
the induced  and piled up magnetic field around  the planetary
obstacle represents an effective  magnetic barrier, preventing free  entrance of
solar wind plasma to the Venusian  ionosphere (Zhang et al., 2007). The  draping
of the interplanetary  magnetic field (IMF),  the accretion of  magnetic flux by
the planet, the characteristic spatial  scales of physical processes as  well as
the  location  and specific  features  of boundaries,  all  depend on  dynamical
processes in the  solar wind and  make this environment  unique. The results  of
Venus Express spacecraft (VEX) provide an altitude for the induced  magnetopause
of about 300  km at the  subsolar point, while  the subsolar bow  shock distance
from the  surface of  the planet  is about  1900 km  (Zhang et al., 2007). These
values exhibit  a solar  cycle dependence,  i.e. the  thickness of  the Venusian
magnetosheath varies around 1500 km at the subsolar point and widens up to  5000
-  7000 km  at the  terminator.
During time intervals of high solar wind  dynamic pressure, the ionopause moves
to low altitude ($\sim 250$ km) and mainly the ionosphere forms the obstacle responsible
for deflecting the solar wind flow. In this case the thickness of the ionopause
increases and more direct interaction between the ionosphere and solar wind is
possible (Elphic et al., 1981; Russel and Vaisberg, 1983).

In  the dayside  magnetosheath strong  magnetic
fluctuations and  waves are  present (e.g.  Luhmann et  al., 1983).  Mirror mode
waves were  observed here  in case  studies (Volwerk  et al.,  2008a), and  also
investigated  statistically (Volwerk  et al.,  2008b). On  the other  hand, the
limited spatial scale of  the magnetosheath on the  dayside might not support  a
development of a turbulent cascade,  the fluctuations rather resemble 1/f  noise
(f is the frequency).  1/f$^{\alpha}$  noise is a  signature of the presence  of
independent  physical  mechanisms  driving  fluctuations  in  the  magnetosheath
(V\"or\"os et al., 2008). Keeping in mind that the ion inertial length is of the
order of 100  km, the magnetohydrodynamic  (MHD) spatial scales  in the Venusian
magnetosheath are limited to one or two decades in wave-number space.

The terminator and, further downstream,  the night-side region is of  particular
interest (Spreiter and Stahara, 1992). Here plasma instabilities, vortices,  and
turbulence can develop  near boundaries. For  example, Wolff et  al. (1980) have
shown that the distortion of the ionopause by Kelvin-Helmholtz instability might
lead to the formation of magnetic 'flux ropes' inside the ionosphere as well  as
ionospheric 'bubbles' embedded in the solar wind. Numerical simulations indicate
that the Kelvin-Helmholtz instability can  occur at the terminator ionopause  of
Venus  (Terada et al., 2002), capable of producing wave structures over 1000  km
in size (Amerstorfer et al., 2007;  Biernat et al., 2007). In fact,  the initial
VEX observation detected these wavy structures (Balikhin et al., 2008). A recent
study shows that MHD turbulence near and immediately after the terminator is not
fully developed because  of the rapid  decrease of spectral  power toward higher
frequencies, resulting in spectral scaling  indices $\alpha > 2$  (V\"or\"os  et
al., 2008).  Further downstream,  in the  magnetotail/magnetosheath region,  the
spectral analysis indicates the presence of developed inertial range turbulence,
with a spectral scaling  index $\alpha \sim 1.6$,  close to the values  expected
for hydrodynamic or  magnetohydrodynamic turbulent flows.  However, in the  same
region,  non-gaussian  probability  density functions  (PDF)  with  typical long
tails, corresponding to intermittent turbulence, were not observed (V\"or\"os et
al., 2008). This  can be  explained by  the shortness  of turbulent  time
series during a tail-crossing, in particular not allowing a full  reconstruction
of PDF tails.

The near-Venus tail is highly structured. There is no significant plasma  inflow
into the near-Venusian  wake. The wake  magnetic field is  known to be  stronger
than the IMF and  from the cavity hot  plasma is excluded (Bauer  et al., 1977).
The  solar   wind-ionosphere  interaction   in  the   presence  of   the  draped
interplanetary magnetic field, however,  produces an extended boundary  tailward
from the terminator, at the inner edge of the magnetopause, or outer edge of the
wake downstream, where  boundary layer turbulence  can develop and  heat locally
the plasma. The width  of this boundary layer  is limited, does not  include the
whole  plasma  sheet. In  order  to understand  the  energy content  and  energy
dissipation   of  underlying   processes,  it   is  necessary   to  investigate
systematically the statistical features  of fluctuations in the  structured near
-Venusian  plasma  environment.  In  this  paper  we  statistically  analyze the
spectral scaling features  of fluctuations in  the magnetosheath, wake  and near
-boundary regions using  VEX magnetometer data  during the first  twenty days in
May 2006. The time resolution of the magnetic data is 1 s. Due to rapid crossing
of  different  structures,  this  time  resolution  is  necessary  for obtaining
statistically reliable scaling results. The main emphasis is on the analysis  of
scalings,  i.e. on  the evaluation  of the  continuous part  of magnetic  power
spectra. The analysis of waves (when the continuous parts of the spectra are not
considered)  is  equally important,  but  out of  the  scope of  this  paper. In
addition to spectral estimations,  we will compare the  probability distribution
functions (PDFs) and evaluate the  scale dependency of their shapes,  associated
with noisy  and turbulent  fluctuations. This  helps to  obtain a  more reliable
differentiation  between   turbulence  and   noise,  occurring   along  the  VEX
trajectory. The near-polar orbit of VEX with a periapsis altitude of 250-350  km
allows for the first time observations at terminator and mid-magnetotail regions
(Zhang  et al.,  2006). These  two important  regions were  not covered  by the
previous missions, e.g. Pioneer Venus Orbiter (PVO, Russell, 1992).

\section{Near-Venus plasma regions with varying spectral scaling properties}  It
was reported by V\"or\"os et al. (2008) that the value of spectral scaling index
$\alpha$, describing the self-similarity of  the power spectrum of the  magnetic
field data in the  frequency range 0.03 -  0.5 Hz, exhibits different  values in
different regions near Venus. The magnetic fluctuations are non-stationary, e.g.
in  the magnetosheath  the magnetic  field strength  is increasing  towards the
induced magnetopause,  which introduces  a trend  into magnetic  field data.  In
order to estimate $\alpha$ robustly, we  used a wavelet method proposed by  Abry
et  al.  (2000) and  applied  it successfully  to  the description  of  magnetic
fluctuations in the Earth's plasma sheet (V\"or\"os et al., 2004). In this paper
we use the Daubechies wavelets, for which finite data size effects are minimized
and  the number  of vanishing  moments can  be changed.  The latter  feature of
Daubechies analyzing wavelets is essential to cancel the influence of polynomial
trends or  periodic structures  in the  data on  the estimation  of the  scaling
index.

Let us demonstrate first  how the spectral scaling  features vary along the  VEX
trajectory  during  its  journey  from  the  dayside  magnetosheath  through the
terminator region and wake to the post-terminator magnetosheath. As an  example,
we show  the variation  of total  magnetic field  strength $B$  on May  19, 2006
(Figure 1 top)  together with the  corresponding power spectral  densities (PSD)
calculated  during   equally  long   time  intervals   (\textbf{a},  \textbf{b},
\textbf{c},  \textbf{d} and  \textbf{e}, Figure  1, bottom  subplots) along  the
spacecraft trajectory.  Zhang et  al. (2007)  have already  demonstrated that  a
crossing  of  each  physical  region in  the  near-Venus  plasma  environment is
recognizable from the  variations of strength  and wave/fluctuation activity  of
the magnetic field. Indeed, $B$ is not disturbed before $t_1=$0115 UT and  after
$t_2=$0310 UT, the  spacecraft is in  the solar wind  (Figure 1). The  planetary
obstacle perturbs the  magnetic field only  between $t_1$ and  $t_2$. During the
interval \textbf{a}  VEX enters  the dayside  magnetosheath from  the solar wind
(VEX trajectories  and the  intervals \textbf{a-e}  are shown  in Figure 2, left
bottom subplot). Here  the magnetic field  strength is strongly  fluctuating and
its value increases  up to $\sim$  50 nT. The  estimated spectral scaling  index
$\alpha = 1 \pm 0.2$ (the first  bottom subplot in Figure 1) is low,  indicating
that fully  developed turbulence  is absent  in this  region (V\"or\"os  et al.,
2008). Higher values $\alpha = $5/3 or 3/2 are expected for hydrodynamic or  MHD
inertial range  turbulence, respectively.  After $\sim$  0135 UT  $B$ decreases,
which corresponds to the closest approach to Venus, VEX is close to or below the
induced magnetopause (Zhang et al. 2007).
%During the twenty crossings of this region in May 2006  scaling indices between
%$\alpha =$ 0.5 and 1.4 were observed
%The bow shock is depicted by the thick gray line and the magnetopause by the dashed black line in Figure 1b.
%The magnetic field strength is strongly fluctuating and its value increases up to $\sim$ 40 nT within the %magnetosheath (interval between 0117 and 0130 UT in Figure 1a). Then $B_T$ decreases almost to zero, which corresponds %to the closest approach to Venus,
%VEX is below the induced magnetopause.
Between 0140 and 0205 UT high frequency fluctuations are absent, only low frequency wavy fluctuations are seen. These wavy structures might be associated with Kelvin-Helmholtz instability at the terminator ionopause or detached plasma clouds near the terminator, observed already during Pioneer Venus orbits (Brace et al., 1982).
During interval \textbf{b} the corresponding spectrum exhibits significant wave power only near 0.07 Hz ($\sim$ 14 s) then the spectral power rapidly decreases with a scaling index $\alpha = 2.5 \pm 0.2$ (the second bottom subplot in Figure 1). Further downstream (after 0205 UT in Figure 1) broad-band fluctuations occur again. The spectral indices within the intervals \textbf{c} and \textbf{e} are $\alpha = 1.5 \pm 0.2$ and $1.6 \pm 0.2$ respectively, indicating the presence of developed turbulence. In contrary, the interval \textbf{d}, in between \textbf{c} and \textbf{e}, shows again 1/f noise-like scaling behavior (the last three bottom subplots in Figure 1).

The physical difference between turbulence and noise is clear.
Turbulence in the wake is a consequence of
nonlinear multi-scale interactions and it is strongly dissipative, heating the background plasma
at the small scales. Noisy fluctuations, exhibiting $1/f^{\alpha}$ scaling behavior with
$\alpha$ around 1, may have multiple physical sources not connected with nonlinear interactions,
typical for turbulence. $0 < \alpha < 1$ over higher frequencies (around 1Hz) can also be associated
with the noise of the magnetometer (V\"or\"os et al., 2004). The low values of
$\alpha$ can also indicate that the spacecraft is not in a physical region where
strong nonlinear interactions and turbulence can exist, e.g., due to low plasma density
or lack of plasma flows. In the following, the notation '1/f noise' refers to 1/f$^{\alpha}$  noise with spectral scaling index in a range of $\alpha \in (0.6, 1.4)$
(find below the definition of the spectral index used in this paper).

In this paper we put the emphasis on statistical examination of the circumstances under which developed turbulence occurs in the post-terminator wake and magnetosheath regions.
\section{Statistical analysis of magnetic fluctuations}

%The horizontal dashed lines in Figure 1a show data intervals with well discernible scaling indices in the Venusian %magnetosheath (1/f$^\alpha$ noise: $\alpha \sim $ 1), terminator (large-scale wavy structures: $\alpha \sim $ 2.5) and %wake (turbulence:  $\alpha \sim $ 1.6, or/and noise: $\alpha \sim $ 1), respectively.
We investigated the time series of
magnetic field strength statistically, obtained in the near-Venusian space during  the
first twenty days in May 2006. During one day one orbit is performed, therefore,
during the  twenty day  interval we  have observed  twenty crossings.
Due to the non-homogeneity and dynamical nature of the near-Venus physical regions the occurrence
and the length of at least quasi-stationary intervals slightly differs each day.
We identified the approximate beginning and end of the steady intervals computing the scaling
indices over time scales 2-30 s, using the wavelet method within sliding overlapping windows.
After that, changing slightly the starting point and the length of turbulent/noisy data,
optimized data intervals with smallest errors in $\alpha$ were found.
We rejected time intervals where the error of the estimation of $\alpha$ was larger than $\pm 0.3$.
The available steady intervals in the dayside magnetosheath are shorter, so we selected 9 min long intervals there.
Further downstream 12 minute long intervals were selected. Selecting longer data
periods would significantly reduce the number of available intervals, while shorter
data sets would decrease the statistical reliability of results.
Data intervals with fluctuating magnetic field shorter than 12 minutes were not included into our analysis.
Due to occasional data gaps or shortness of steady fluctuations in the time series, the number of events
is not the same in different physical regions.
There were crossings with no steady fluctuations along the VEX trajectory.
There were also crossings
where the time interval with steady fluctuations was longer than 24 minute.

Due to data gaps or shortness of statistically stationary time series,
only 17 out of 20 crossings are shown in Figure 2 (left bottom and right subplots).
A cylindrical coordinate system is used in Figure 2, where the events are shown in $\surd(Y^2+Z^2)$ VSO vs. $X$ VSO (bottom left) or in VSO coordinate pairs (right subplots).
VSO is the Venus-centered Venus-Sun-Orbital coordinate
system, where X is in the direction of the Sun, Y opposite to the orbital direction of Venus
and Z perpendicular to the orbital plane, positive to ecliptic north.

The near-Venusian space
is physically non-homogeneous and the spatial and temporal variations cannot be
straightforwardly separated  from single-spacecraft  measurements.
Nevertheless, during the investigated time period in May 2006,
the typical sequence of physical regions visited by  VEX
remained  approximately  the same  as  in Figure  1:  crossing of  the  dayside
magnetosheath  (increasing  $B$, interval \textbf{a}),
low-frequency  wavy  structures
after the terminator (interval \textbf{b})
and entering into the region of broad-band fluctuations further downstream
(intervals \textbf{c-e}). The intervals \textbf{a-e} were introduced for the event
on May 19, 2006 (Figure 1: top and Figure 2: top-left).
The wavy structures are not always present along the whole near-terminator region and wake. In the absence of the wavy structures
only low amplitude magnetometer noise is observed. Possibly, the occurrence/absence of wavy structures
can be associated with the upstream conditions. For example, during times when the solar wind dynamic pressure is high
the ionopause moves to low altitudes ($\sim$ 250 km), and direct interactions between the ionosphere and the solar wind
can occur (plasma-plasma interactions or solar wind electric field - ionospheric currents interactions). When the solar wind
dynamic pressure is low the ionospheric width increases (Elphic et al., 1981) and the draped IMF above $\sim 300$ km can stop the solar wind more
efficiently. In this paper we investigate only the features of magnetic fluctuations, without considering the changes in the upstream conditions.
The open triangles (Figure 2: bottom left) indicate the whole region (marks are only on the lowest trajectory)
where the wavy structures appear during the first 20 days in May 2006.

Due to the similarity of crossings,
the depicted intervals (lines) along the trajectories (Figure 2: bottom-left), in terms of spectral properties, refer
generally to similar physical regions along the VEX trajectory. For example, the thick black lines
(Figure 2: bottom-left) indicate turbulent processes identified through a spectral scaling index
near $\alpha = 1.6$ . The location of turbulence in space approximately coincides with
time intervals \textbf{c} and \textbf{e}.
The time interval \textbf{d} coincides roughly
with the position of grey '+' signs along the trajectory, where scaling indices corresponding to $1/f^\alpha$ noise were observed.
For a better visibility, the trajectories together with turbulent and noisy time intervals (\textbf{c}, \textbf{e} and \textbf{d})
are depicted in the $VSO$ coordinates (Figure 2: right subplots).
From Figure 2 (right bottom) it is visible that the spatial regions exhibiting the same
$\alpha$ along multiple trajectories are partially overlapping, indicating that these regions and the
corresponding boundaries are moving or the fluctuations are patchy or intermittent.
Moving boundaries can appear in connection with changing conditions in the solar wind.

Switching between temporal and spatial coordinates helps us to identify spatial regions with typical scalings.
Occasionally we will use plural indicating the change between temporal and spatial coordinates. For example,
the notation 'intervals \textbf{c}' means the set of all crossings in space near the time interval/space region \textbf{c}
on 19 May, 2006.

Figure 3 shows histograms with the number of events per unit interval
of the scaling index $\alpha$, estimated during the time periods \textbf{a}-\textbf{e}.
In each subplot, the horizontal lines over the histograms represent the average of 95 $\%$
confidence limits for the mean in each value of $\alpha$. The largest average uncertainty corresponds
to the dayside magnetosheath region, where the length of analyzed data was shorter.
Figure 3 indicates that the specific scaling features characterize the statistical properties of
different regions in space rather well.

During the magnetosheath periods (intervals \textbf{a} and \textbf{d}: top and bottom subplots in Figure 3) only two events display scaling indices $>$ 1.3, for the other events $\alpha \in (0.6 - 1.25)$. This indicates that the fluctuations or waves
present in the magnetosheath, except of a few cases, usually do not evolve to a fully developed turbulence state. The observed range of scaling indices suggests that the continuous part of the magnetosheath spectra is associated
with $1/f$ noise rather than turbulence. Noisy fluctuations can be convected from the dayside magnetosheath
(interval \textbf{a}) to the post-terminator magnetosheath downstream (interval \textbf{d}),
where the spectral power is smaller, possibly the driving sources are more distant than at the dayside
(compare the spectra corresponding to the intervals \textbf{a} and \textbf{d} in Figure 1).

In the post-terminator wake (intervals \textbf{b} in Figure 3) wavy structures dominate with periods from 5 to 50 s (similar to event \textbf{b} in Figure 1). Since towards higher frequencies the power rapidly decreases ($\alpha \sim 2.5$) independent driving sources associated with broad-band noise or nonlinear multi-scale turbulence are absent.

Finally, a well discernible turbulence scaling index $\alpha \sim 1.6$ was observed
during the intervals \textbf{c} and \textbf{e} (Figure 3: third subplot from top).
The events from both intervals are plotted together because the corresponding scaling indices are
similar.

The cartoon in Figure 4 helps to interpret the occurrence of typical scaling regimes, summarizing the results of the statistical analysis
along VEX trajectory in the cylindrical VSO coordinate system: (1) 1/f noise
appears under rather different conditions in the magnetosheath: in front of the planetary obstacle
and downstream in the post-terminator magnetosheath (regions \textbf{a, d}: squares in Figure 4). Scaling indices which can be associated with turbulence
occurred in $\sim 5 \%$ of cases (see Figure 3); (2) Coherent wavy fluctuations occur between the dayside magnetopause and
post-terminator boundary of the wake (largely within the optical shadow, region \textbf{b}: triangles in Figure 4),
where noise and turbulence are absent (only magnetometer noise is present when the wave activity is absent).
Closer to the terminator, the wavy structures can be associated
with  the Kelvin-Helmholtz instability occurring at the terminator ionopause, resulting probably in detached plasma clouds,
first observed during the Pioneer Venus mission by Brace et al. (1982). The PVO spacecraft observed filamentary structures (rays),
density holes and radially aligned draped magnetic field lines in the near-planet night side wake (Luhmann and Russel, 1983; Marubashi et al., 1985);
(3) Developed turbulence is present at or near boundaries: at the wake/magnetopause/magnetosheath boundary and near the bow shock (regions \textbf{c, e}:
circles in Figure 4).

\section{Discrimination between boundary turbulence and magnetosheath noise}
Figure 3 shows that, in terms of spectral scaling characteristics, turbulence
and noisy intervals are well separable.
In the following we will further investigate some other differences between turbulence (intervals \textbf{c} and \textbf{e}) and noisy magnetosheath fluctuations (intervals \textbf{d}).
We first recall some similarities between the observations obtained from previous missions and VEX near the regions with developed turbulence.

Early observations from Mariner 5 (Bridge et al., 1967)
already showed the existence of a plasma boundary located at the inner edge of the post-terminator magnetosheath
(VEX is crossing this region where the occurrence of turbulence is depicted by the bottom circle in Figure 4).
During a flyby, Mariner 5 found the location of the boundary layer from $\sim 3.2$ to 4.5 $R_V$ behind Venus and from  $\sim 2$
(magnetopause location) to 3.5 $R_V$ from the center of the wake.
Within the boundary layer Mariner 5 observed a strongly fluctuating lower intensity magnetic field, a
decrease of the density and velocity, but an increase of the temperature.
The magnetic field strength was larger when the spacecraft entered from the inner magnetosheath,
across the magnetopause, to the wake.
VEX observations show that the magnetopause location is rather variable over
distances $X VSO < -1 R_V$ (open circles in Figure 2: left bottom). Therefore,
multi-point turbulence fluctuations indicating the crossing of the magnetopause start in
different distances from the center of the wake. For example, on May 19, 2006 (Figure 2: left top)
the strongly fluctuating magnetic field strength reaches a local maximum at about $ 0205$ UT, then it slowly decreases.
It indicates that VEX is entering to the boundary layer from the wake (from intervals \textbf{b} to \textbf{c})
and at the boundary the character of fluctuations is suddenly changing, from scaling index $\alpha \sim 2.5$ to $\alpha \sim 1.6$
(see also Figure 1).

Another interesting feature includes the change of the sign of $B_X$.
During the interval \textbf{c} on May 19 it happens several times (Figure 2: top left).
It can indicate that VEX is crossing the neutral sheet behind the planet.
The VEX trajectory for this day (marked by an arrow in Figure 2, top right subplot) is roughly along Y VSO $\sim 0.9 R_V$.
In fact the Venusian wake/magnetotail  can be formed  by flux tubes  convected around or
slipping over the planet  and filled with the  plasma from the day  side. At the
same time, both ends of the flux  tubes are co-moving with the solar wind.  In a
simple case, the resulting magnetic field  in the tail is stretched along  the X
axis, because  the central  part of  the flux  tubes is  slowed down by the planet
while both  ends travel  faster with  the solar  wind. Since  the IMF is usually
close to the ecliptic  plane, the tail current  sheet, in ideal case,  should be
formed by the stretched magnetic field mainly in the north-south plane  (Russell
and Vaisberg, 1983). Depending  on the upstream IMF  and its draping around  the
planet, the  current sheet  plane in  the Venusian  magnetotail can  also rotate
(Luhmann et al., 1991). Since turbulence can drive local mixing and vorticity, the change of the magnetic field direction
might be also associated with turbulence, not necessarily with neutral sheet crossing.
For example, $B_X$ is also changing sign several times during the shock associated turbulence
interval \textbf{e} (Figure 2: top right), but it has nothing common with a neutral sheet crossing.

The same magnetic field and plasma signatures of the inner magnetosheath boundary layer were observed by other missions, too.
For example, Venera 10 observed the boundary layer tailward from the Mariner 5 flyby (Romanov et al., 1978).
The near-terminator part of the boundary layer was investigated during the Pioneer Venus Orbiter mission.
Perez-de-Tejada et al. (1991, 1993)  have seen the boundary between the bow shock
and magnetopause (they are using the term ionopause instead) in the vicinity of and downstream from the terminator.
The boundary layer observed in different distances along the inner edge of magnetosheath
confirms the persistent presence of this boundary as a rarefaction wave emerging from the terminator magnetopause
and extending downstream.
The high level of magnetic fluctuations and occurrence of magnetohydrodynamic waves near a quasi-parallel Venusian bow shock is well-known
from the PVO  mission (Luhmann et al., 1983).
In this paper we emphasize the detection of intermittency near the Venusian bow shock and its differentiation from
magnetosheath noisy fluctuations.
Also, we investigate only the magnetic field signatures of the boundary layers.

\subsection{Magnetic field orientation}
First, we  investigate the
local magnetic conditions under which turbulence (or noise) can appear.
For this purpose we calculate the instantaneous contribution of magnetic
field  VSO components to the total magnetic field:
$b_X(t)=B_X(t) / B(t)$, $b_Y(t)=B_Y(t) / B(t)$ and $b_Z(t)=B_Z(t) / B(t)$.
In each time step $t$ the sum $b_{X}^2(t)+b_{Y}^2(t)+ b_{Z}^2(t) = 1$.
Our goal is to compare the average magnetic field direction during turbulent (\textbf{c, e}) and noisy (\textbf{d}) intervals, when the scaling index is well defined (see Figure 2). Since $\alpha$ is a statistical descriptor over
12 min long intervals the mean values $<b_{X}^2>$, $<b_{Y}^2>$, and $<b_{Z}^2>$, respectively, were
computed over the same time periods; their sum is again 1. Each average defines the relative contribution
of a magnetic VSO component to the average magnetic strength during the considered intervals.

Figure  5  compares the  local  mean magnetic  field  components with  the
values of  scaling indices estimated  during the
same  12 min  long intervals (open and filled circles).
The horizontal lines are averages of points  over the
noisy $\alpha < 1.4$ and turbulent $\alpha = 1.6 \pm 0.2$ scaling index ranges.
These distinct ranges of $\alpha$ correspond to the statistical results in Figure 3.
The filled circles correspond to turbulent intervals within the magnetosheath boundary layer, while, over the same range of $\alpha$s, the open circles correspond to bow shock associated turbulent intervals.

The horizontal lines in Figure 5 show that, over the turbulence range, the average of $<b_{X}^2>$ and $<b_{Z}^2>$ is larger
while the average of $<b_{Y}^2>$ is smaller than over the noise range of scaling indices.
Therefore, when turbulence is observed  $<b_{X}^2>$ dominates over $<b_{Y}^2>$ and the latter is still
stronger than the increased $<b_{Z}^2>$. Magnetosheath boundary layer associated turbulence intervals
(filled circles - intervals \textbf{c} in Figure 2) show even larger $<b_{X}^2>$ values than the near-shock turbulence values (open circles - intervals \textbf{e} in Figure 2,
over the same $\alpha$ range). In accordance with our findings, the analysis of PVO magnetic, electric and plasma data has shown that, near the terminator region but within the magnetosheath boundary layer, the magnetic field is nearly aligned with
the Sun-Venus line (X VSO direction). It was interpreted in terms of a friction-like viscous interaction between the shocked solar wind and the ionospheric plasma over the magnetic polar regions where the draped interplanetary magnetic field lines slip over the planet (Perez-de-Tejada et al., 1993). The viscous plasma-plasma interaction and frictional heating can explain the enhanced temperatures inside the
magnetosheath boundary layer, observed tailward by Venera 10 (Romanov et al. 1978). An alternative explanation is represented by local turbulent heating of the boundary layer plasma.
We note that near the magnetopause, where VEX crosses the turbulent boundary layer, the draped interplanetary magnetic field lines are more stretched having a large X VSO component (Figure 4).
Although the number of events is rather limited, the near-shock turbulent regions have smaller X VSO but larger perpendicular
Y VSO and Z VSO magnetic components than the magnetosheath boundary layer turbulence (Figure 5).

Finally, noise (open circles - intervals \textbf{d} in Figure 2) is associated
with a large $<b_{Y}^2>$, smaller $<b_{X}^2>$ and almost negligible $<b_{Z}^2>$.
Since the IMF is mostly in the X-Y plane, the large-scale magnetosheath magnetic field is not affected by the boundaries, but exhibiting only noisy broad-band fluctuations and it has the strongest average components in the X-Y plane.

Let us investigate deeper now the intermittent  nature of
turbulence. Besides the characteristic scaling exponents ($\alpha \sim 1.6$) and the expected average magnetic field directions,
intermittency represents another key feature of fully developed turbulence.
Therefore, the occurrence of intermittency represents a further evidence that we are
dealing with a real turbulence, capable of heating the background plasma.

\subsection{Turbulent intermittency versus Gaussianly distributed noise}
Higher order statistics is needed to fully describe the nature of nonlinear fluctuations. In turbulent,  non-homogeneous plasma flows the shape of the probability density functions (PDFs) is scale-dependent and  peaked with long tails (e.g. Frisch, 1995). Because of the shortness of time series  during one crossing the shapes of non-Gaussian PDFs or their scale dependency cannot be evaluated (V\"or\"os et al., 2008).  Instead, we construct PDFs from the 15 magnetic time series (realizations) of turbulence for which
spectral scaling near $\alpha \sim 1.6$ was observed (events from both \textbf{c} and \textbf{e} intervals in Figure 2). Splitting the data
into magnetosheath boundary layer and near-shock turbulence regions would not change the shape of turbulent PDFs significantly. PDFs corresponding
to the noisy magnetosheath (intervals \textbf{d} in Figure 2) will be also reconstructed.

PDFs of two-point differences of magnetic field strength were estimated
from  $\delta B  = B(t+\tau)-B(t)$,  for $\tau  = 2...30$  s. Figure 6 shows the PDFs  for
$\tau  = 3, 30$  s only. Boundary turbulence associated PDFs are shown on the left, 1/f noise related PDFs on the right hand side. The error bars represent 95\% confidence limits for the mean in each point of
$\delta B$. The dashed grey points are least-square Gaussian fits to the experimental two-point PDFs.
The Gaussian PDF is  given by $\frac{1}{\sigma \surd (2\pi )}\exp ( -\frac{(x-\mu )^2}{2\sigma ^2} )$,
where $\mu $ is the mean and $\sigma ^2$ is the variance. For the smaller time scale $\tau=3$ s the tails of the experimental turbulent PDFs are higher than the Gaussian tails. The departure from the Gaussian indicates that non-homogeneously
distributed fluctuations become more probable as the scale decreases due to turbulent structures and long-range
interactions (Leubner and V\"or\"os, 2005). Gradual decorrelation is obtained by enhancing the two-point separation scale and a Gaussian is approached for large enough $\tau$ even in a turbulent field. This is because the typical correlations for turbulent structures are lost if the separation is large, and only Gaussianly distributed noise is observed. For $\tau=30$ s, the PDF is a Gaussian in Figure 6 left.
Noise shows Gaussianly distributed PDFs over both time scales.

Let us further investigate the shape of PDFs in terms of statistical moments.
Using standard procedures (Press et al., 1992), the skewness ($S$)
or the third moment,
\begin{equation}
S(\tau)=\frac{1}{N}\sum_{j=1}^{N}\Big\lgroup \frac{x_j-\overline{x}}{\sigma}\Big\rgroup^3
\end{equation}
and the kurtosis ($K$) or the fourth moment,
\begin{equation}
K(\tau)=\frac{1}{N}\sum_{j=1}^{N}\Big\lgroup \frac{x_j-\overline{x}}{\sigma}\Big\rgroup^4 - 3
\end{equation}
are computed for turbulent and noisy intervals, as above.
Here $x_j=\delta B(t_j, \tau)$, $\sigma$ is the standard deviation, $\overline{x}$ is the mean
value of the elements and $N$ is the number of the data points.
In this way  the dimensionless $S(\tau)$ and $K(\tau)$
characterize the scale and time evolution of the shape of a distribution
($S$ - asymmetry around the mean; $K$ - peakedness or flatness; both relative to a Gaussian distribution,
for which $S=K=0$).
$K$ increases towards small scales in intermittent turbulence (Frisch, 1995).

Figure 7 shows the time-scale ($\tau$) evolution of kurtosis $K(\tau)$ (top) and skewness $S(\tau)$ (bottom) for turbulent (left) and noisy (right) intervals.
The error bars represent 95\% confidence limits for the mean in each value of $\tau$.
As is expected from turbulent PDFs in Figure 6, $K$ is increasing as $\tau$ decreases, which proves that
the underlying magnetic fluctuations are non-Gaussian, peaked and intermittent.
$K$ practically does not depend on $\tau$ in the case of 1/f noise, indicating a Gaussianly distributed process.

The skewness remains close to zero (Figure 7: bottom) in both cases, showing symmetric distributions around the mean value.

\section{Discussion and conclusions}
In this paper the unique data  from the VEX spacecraft were used  to investigate
magnetic fluctuation statistics from   Venusian magnetosheath and wake  regions.
The interaction of the solar  wind with the planet drives  magnetic fluctuations
exhibiting  different  scaling  regimes in  different  regions  of near-Venusian
space.  To identify  spectral scaling  ranges and  indices, we  used a  wavelet
technique,  successfully  applied for  studying  the continuous  spectra  in the
Earth's plasma sheet turbulence (V\"or\"os et al., 2004). The technique and  the
data intervals were not optimized for finding waves (peaks in power spectra).

Three types of scaling were observed. Inside the dayside/tailward magnetosheath,
far from boundaries, 1/f  noise is present in  the prevailing majority of  cases
indicating the  contribution of  multiple independent  driving sources.  It also
means, that this type  of spectral scaling is  not formed by an  isolated single
physical  process.  In  other  words,  there  might  exist  multiple  sources of
fluctuations, but no  one of them  is close enough  in space to  dominate in the
spectral power. This changes when VEX enters to a region where multiple  sources
are missing, or where a single physical process dominates with a scaling feature
other  than that  of the  noise. In  fact, these  are distinct  regions of  near
-Venusian space  where the  'solar wind  - planetary  obstacle' interactions are
enhanced or the multiple sources of noise are shielded.

The  interaction is  enhanced at  the terminator  ionopause, probably due  to the
Kelvin-Helmholtz  instability,  at  the  magnetosheath  boundary  layer  due  to   the
magnetosheath boundary shear  flows and near  the quasi-parallel bow-shock.  The
near-planet wake represents a region which is shielded from the plasma of  solar
wind origin.  Filamentary structures,  detached plasma  clouds, depleted density
holes and  radially aligned  draped magnetic  field lines  were observed  by PVO
spacecraft  in  the  near-planet  night side  wake  (Luhmann  and  Russel, 1983;
Marubashi et al., 1985). The magnetosheath flow is expected to converge into the
wake  only near  5 $R_V$  behind the  planet (Intriligator  et al.,  1979). The
observed wavy structures near the  terminator and in the night  side near-planet
wake can be associated  with the detached coherent  structures or holes.
The occurrence/absence of these structures can be controlled by the direct interaction between
the solar wind and ionosphere, e.g. by the high/low solar wind dynamic pressure.
In  our
interpretation,  the  spectral  index  $\alpha\sim  2.5  $  indicates,  that the
coherent  wavy  structures represent  the  dominating physical  process  in this
region. Because of the  shielding of the near  planet wake, turbulence or  noise
are  absent  in this  region.  Due to  the  converging flows  the  shielding can
disappear at distances  close to or  larger than X  VSO $\sim 5  R_V$, where the
character of fluctuations would change.

The magnetosheath regions  with distinct scaling  indices (turbulence or  noise)
partially overlap (see Figure  2). This can be  explained through a movement  of
boundaries  under the  influence of  changing upstream  IMF conditions.  Spatial
intermittency,  typical   for  turbulence   with  scale-dependent   non-Gaussian
distributions, can  lead also  to interwoven  scaling structures.  The turbulent
regions are formed  near the 'supersonic  solar wind flow  - planetary obstacle'
boundaries in the presence of draped  IMF. The outer boundary is the  bow shock,
where the solar wind slows down and  heated for the first time. Here, the  local
turbulence is associated with the quasi-parallel shock geometry. Second time the
solar wind  flow decelerates  at the  inner magnetosheath  producing a  velocity
shear near the magnetosheath boundary layer. The near-terminator ionosphere/magnetopause  and
its interaction  with the  solar wind  plays an  important role. The rarefaction
wave (the boundary) and the observed plasma conditions in the tailward  boundary
layer can emerge from the magnetopause near the terminator and extend downstream
(Perez-de-Tejada et al., 1991).  Local heating of the plasma within the boundary
layers  is  also possible  through  the shear  flow  associated turbulence.  The
spatial size of boundary layer turbulence (see Figure 2) is roughly between  0.5
and 1 $R_V$, shorter turbulent intervals were observed near the quasi-parallel bow  shock.
The width of the noisy magnetosheath  in between the turbulent boundaries is  of
the same order. The  estimation of the spatial sizes of these regions is rather rough.
The data are available only  from single  point measurements  (the boundaries  can move
during the measurements) and the 12 min long data intervals represent a pure resolution.  The
estimation of  scaling indices  within shorter time intervals, however, was not  possible due  to the  large statistical
errors in these cases.
Multi-point Cluster observations show, that turbulence downstream of the
quasi-parallel terrestrial bow shock is intermittent and that the level of intermittency increases over
the spacecraft separation, reaching larger values than 8000 km (Yordanova et al., 2008).
Our results show, that the spatial scale of intermittent turbulence is less than 1 $R_V$
near the Venusian bow shock.
In between the near shock region and the magnetosheath boundary layer fluctuations are noisy (in Figure 4, squares between triangles).

It was shown by Perez-de-Tejada et al. (1993) that the magnetic field is  nearly
aligned with  the Sun-Venus  line (X  VSO direction)  within the near-terminator
magnetosheath boundary layer. We found  a similar alignment within the  boundary
layer at a distance of X VSO $\sim -2.2 R_V$ (Figures 2, 5). It indicates,  that
the  magnetic   field  geometry   detected  during   the  viscous  plasma-plasma
interactions near the  terminator ionosphere is  conserved and observed  further
downstream along  the VEX  trajectory, where  the draped  magnetic field is
stretched (Figure 4).

Data  intervals  that were found to be turbulent  or  noisy in the  spectral  analysis were  further
investigated using  the two-point  (time delayed)  probability density functions
(PDFs). Over large two-point separations (e.g. $\tau=30 s$), both turbulent  and
noisy  PDFs  are  well  fitted by  the  Gaussian  distributions,  indicating the
occurrence of uncorrelated fluctuations.  Over small separations (e.g.  $\tau= 3
s$),  large deviations  from the  Gaussian distribution  are observed  only for
turbulent intervals, noise remains Gaussianly distributed (Figure 6). The  scale
-dependency  of  kurtosis  (Figure  7)  shows  that  turbulent  structures   are
intermittently distributed, noise is more homogeneous. Skewness remains close to
zero  in  both   cases  which  corresponds   to  symmetrical  distributions.   A
simultaneous increase of  $S$ and $K$  towards small scales  would indicate that
the multi-scale  fluctuations in  turbulence might  be affected  by strong large
-scale  gradients  or  close boundaries  (V\"or\"os  et  al., 2007).  Therefore,
$S(\tau) \sim 0$ for  $\tau \in (2-30)$ s  is a signature of  intermittency, not
affected  by boundaries  along the  VEX trajectory  over the  scale of  seconds.
However, asymmetries or nonzero skewness in magnetic field statistics can appear
below the 1 s time scale.  Anyhow, besides the expected spectral index  ($\alpha
\sim 1.6$), the observation of  intermittency represents a further evidence  for
the occurrence of real turbulence in the near-Venusian space. The key feature of
turbulence is its strong dissipative nature  and a capability for the local heating of plasma.
Multi-scale turbulence can channel the large-scale energy of the flow to kinetic
scales, where dissipation processes are  strong. We will investigate this  point
in a different paper.

%% ------------------------------------------------------------------------ %%
%
%  ACKNOWLEDGMENTS
%
%% ------------------------------------------------------------------------ %%
\begin{acknowledgments}
The work of Z.V. and M.P.L. was supported by the Austrian Wissenschaftsfonds under grant number
P20131-N16.
\end{acknowledgments}

%% ------------------------------------------------------------------------ %%
%
%  REFERENCE LIST AND TEXT CITATIONS
%
%% ------------------------------------------------------------------------ %%
%
% If you use BiBTeX for your References, please do not send
% your bibliography database. Copy the reference list
% from your .bbl file into your article file before submission:
%
%1. Run LaTeX on your LaTeX file.
%
%2. Run BiBTeX on your LaTeX file.
%
%3. Open the new .bbl file containing the reference list and
%copy all the contents into your LaTeX file after the
%acknowledgments section;
%
%4. Comment out the old \bibliographystyle and \bibliography commands.
%
%5. Run LaTeX on your new file before submitting.

%Failure to follow these instructions will require manual
%intervention through hard keying of information,
%which can introduce errors.

\end{article}

\end{document}